\documentclass[11pt]{amsart}
\usepackage{color}
\usepackage
           {graphics}
\definecolor{mylinkcolor}{rgb}{0,0.55,0.55}
\definecolor{myurlcolor}{rgb}{0.9569,0.2510,0.0196}
\definecolor{myanchorcolor}{rgb}{0,0,1}

\usepackage[pdftitle={A title},
            colorlinks=true,
            urlcolor=myurlcolor,
            linkcolor=mylinkcolor,
            anchorcolor=myanchorcolor
]{hyperref}

\usepackage{gil}

\title[Space-time databases modeling global semantics networks]%
{Space-time databases modeling global semantics networks}
\date{29.03.2005}
\author{Prikhod'ko A.\,A.}
\author{Prikhod'ko N.\,A.}

\begin{document}

\begin{abstract}
This paper represents an approach to creating global knowledge systems,
using new philosophy and infrastructure of global distributed semantic network
(frame knowledge representation system, FRS) based on
the \termlink{spacetime@database}{space-time database} construction.
The main idea of space-time database environment introduced in the paper
is to bind a \term{}{document} (an information frame, a knowledge)
to a special kind of entity, that we call \termlink{permanent@entity}{permanent entity}, ---
an object without history and evolution, described by a ``point''
in the generalized, informational space-time
(not an evolving object in the real space having history).
For documents (information) it means that document content is unchangeable,
and documents are absolutely persistent.
This approach leads to new knowledge representation and retreival techniques.
We discuss the way of applying the concept to building a global distributed scientific
library and scientific workspace. Some practical aspects of the work are
elaborated by the open IT project \url{http://sourceforge.net/gil/}.
\end{abstract}

\maketitle

\section*{Introduction}

This paper represents an approach to creating global knowledge systems,
using new philosophy and infrastructure of global distributed semantic network
(frame knowledge representation system) based on
the \termlink{space-time@database}{space-time database} construction.
We start our investigation with thinking of universal world wide library
of scientific knowledge, interpreted mainly as a storage of documents in usual human sense.
It is simple to imagine but hard to realize, and
technological complexity is just one side of the problem.
Today a lot of theories and technologies proposing a way of
representing knowledge are drawn up and elaborated,
most of them focus on concrete problems of information processing
and artificial intellegence and few ones pretend to reform
the practice of scientific research and to build a universal
scientific workspace for globalized scientific investigations.
Since our concept closely connected to the problem of scientific
knowledge representation, we emphasize the phylosophical and
humanitarian aspects which allow to see the problem of
knowledge birth and evolution from completely different point of view.

The concepts of \termlink{generic@document}{document} and \termlink{library}{library}
as they used in the paper
are formally independent but actually develop the concept of
\term{}{semantic network} or, in different terminology,
\term{}{frame knowledge representation system (FRS)},
and sometimes inspired by it.

The fundamental idea of semantic network (\see cite{Minsky1})
is to create abstract knowledge systems
composed by concepts and statements that can be applied to
situations of different kind concerning different subjects.
Frame knowledge representation systems (FRS) are commonly based
on the principle of representing concepts and objects as
frames which are entities of special kind joined in a network (graph).
Links or relations between nodes symbolize different mutual relations:
inheritance, aggegation and so on. As it is noted by Karp
in paper \cite{Karp1} containing analysis of more than 50 branches of
semantic network and knowledge representation theory,
there is a variety of terminologies to be specially compared,
so we will use that paper of Karp for a common language to speak
about the matter.

%

The concept of \term{global@scientific@library}{global scientific library} proposed in the paper
is slightly aside the main stream of semantic network theory.
%
%
While the frame knowledge theory considers frames and network
from abstract point of view, we focus on implementation part of the
problem, trying to find a way of giving simple and efficient
material form for distributed knowledge systems to be used in
systems of universal (free world wide interdisciplinary) knowledge access
and global framework for scientific investigations. In some sense
we would like to propose an approach to taking step from particular
and abstract semantic systems to persisten global network similar to
step from programming language to GRID infrastructure, but concerning
knowledge representation systems.


In this paper we 
consider computer document and knowledge representation,
however the entire knowledge of mankind is formed by different types
of information media, huge part of it are the hard copies of books and articles.
Today owing to the develpment of computer technologies and telecommunications
the \term{}{book} and the \term{}{letter} (to wide extend) are not
the unique information media.
Now information takes especial significance.
Concept of information itself is evolving and,
what is important, not only for special investigations but in mind
of an ordinary man (good example is science fiction).
Systems are produced specially for handling information  and processes
over information. New languages for representing information appear.

Now there exists huge variety of technologies of
data and knowledge representation, which are usually complex or
too much specialized.
The problem of knowledge representation has different levels
according to the extent of information technologies integration to
a concrete sciense and its specific needs.
Bioinformatics uses large data networks for a long time,
it demonstrates the high level.
At the same time people make sience without special computer programs
or networking technologies so far. This case is related to the low
level of the problem concerning common ways of representing
scientific knowledge on a universal basis. Our work concerns
this second part of the problem. Simple and sufficiently universal
technologies are still to be discovered to get real advantage
in using global networks. We guess that an important part of
understanding must refer to or include general phylisophical
analysis of knowledge phenomenon considering it in some cases
as natural (physical) phenomenon (but not as a tool confined by
use and tradition!).
%
%
Our investigation will follow this principle. An ideal goal would be
both developing new knowledge management systems and creating new
international scientific language which is simple, semantic-oriented
and self-organized to be universally applied.
%

Chalenging aspect of knowledge phenomenon problem is information self-organization.
Knowledge produced by scientific community grows exponentially, it is hard to put it into
single mind and to analyze using conventional methods base on hand-make analysis.
Today researchers work on new types of knowledge bases with high level of self-organization
resembling biological or gene constructions.

We are interesting in a close but slightly different side of the question:
\emph{}{knowledge automony}, which means investigating simple and general
\emph{knowledge meta-model} serving as a basis for the knowledge to \term{}{exist} independendently
(to be explained in the paper).
Concerning data autonomity our concept should operate the meta-concept,
that we call \term{}{global}, but not \term{}{distributed} (\dots network, knowledge system etc.).
A distributed system is generically considered as whole system represented by plenty of equal
particles. We consider elements of the global knowledge base to be much more independent
both from semantical and operational point of view.

Consider two examples. First, look at a library inside library network. Information in
the library must keep sense if network is disconnected. Futhermore, information must be
independent from ways of representation and media types, --- knowledge can be printed on
paper (serialized) without loss of sense. Semantics must be unchangable during a long period
of time, thousands and millions of years. Thus, an information unit (a library) must have
something special (close to concept of meta-data) ``gluing'' different semantic layers
helping a reader to understand the knowledge. Notice that semantics cannot be made completely
reader-dependable. This paradox seems to be crucial, but the thesis is explained in next example.

Consider another example of self-organization. Suppose that a library as a part of the knowledge
is put on analysis by a Man from another epoch, say, after million of years. We assume that
he has no dirrect connection to our culture. Library is found as a result of archeological dig.
Modern specialists do deep analysis of texts produced by different cultures trying to
guess (conjecture) the sense. In case of art and fiction delivering feeling and emotions
of the writer, of course, one should know well culture of the writer to compare and
track down common sense in our culture. However if we try to reconstruct
a description of a mechanism or a complex technological system, this is not the case.
We deal with a knowledge which is simple and rigorous, composed of elementary parts
in conformity with strict rules and principles. Scientific community developed
non-trivial culture of knowledge representation including structuring and using
specialized languages. Any library must incorporate this experience on way or another
to represent the knowledge of mankind to be understandable and useful
to any human or non-human reader.

The idea of universal basis for common scientific language is to choose as
elements for the knowledge representation system few simple (primitive)
constructions intelligible to any living creature having
\emph{structural way of thinking} and then to extend it
consecutively by advanced \emph{semantic layers} to get complex
constructions.

Even if we think of the initial step of library recognition made by an
imaginary creature-reader introduced above, the use of extremely
simple basic constructions helps to do futher semantic analysis
more effectively, moreover, now it is possible to handle \emph{large}
amount of information in the library and to involve (computer) automation.
Consider the following case. In near fiture the mankind probably will start
exploration of near cosmos, say, planets in solar system. Imagine
an isolated station on orbit or a planet, having in its kernel
a kind of a library discussed above able to self-organize and self-describe.
Any man visiting station after a long period of autonomy
(observe that human culture can change completely in hundreds of years)
will be able easily communicate the station information system
without any deep analysis just having a base constant list presented
on a couple of sheets of paper.

Summarizing the paper approaches to discovering a model of universal
knowledge representation and to providing the low-level environment
for knowledge representation that can be used in high-level knowledge
representation systems. Futhermore, the knowledge environment discussed
in the paper as a kind of distributed database environment explores
new ways of organizing the information as regards: data representation
and exchange, tracking versions and providing meta-data. We hope this
part of the work will be interesting to people working on distributed
information system technologies.

Reader can refer to \cite{Minsky1}, \cite{Karp1}, \cite{Rocha1}, \cite{ARP},
to \cite{OKBC2}, \cite{GFP2}, \cite{KIF}, \cite{CKML}, \cite{Giacomo}
for the knowledge representation theory, IT standards and sample applications.

\section{Global scientific library}

\begin{defn}
Speaking of the \term{generic@global@scientific@library}{global scientific library}
we mean the world wide scientific environment consolidating investigations
belonging to different sciences and schools with help of
global networked knowledge exchange system%
\footnote{As well as common phylosophy of how the Knowledge is organized.}.
In other words, the notion of
the \termlink{global@scientific@library}{global scientific library}
includes a set of concepts: information starage, network,
basis for integrating geterogeneous research systems and knowledge bases,
scientific workplace and so on.
\end{defn}

\begin{defn}
A \term{library}{library} is a specially desgined storage of information
in the wide sense: books, articles (including their computer representation),
algorithms etc.
\end{defn}

Today a \termlink{library}{library}
is not only a storage of hard-copied books, it can be
a library server or just a directory of home computer.
Generally we can consider all the knowledge in the world
(contained in separate libraries)
as the one (global) world library. There is certain analogy with
the Internet connecting workstations and stations.

In most cases it makes not matter for reader (library user)
which library has provided the desired information to him.
Thus, a \termlink{library}{library} providing universal access to the entire knowledge
must be treates as one integral object.
From this global point of view the nodes of the global library
network differ only by information completeness and accessibility
(including time necessary to get some information).
%

\begin{defn}
We call the \term{global@library}{global library} the entire set of
all knowledge in the world (or universe). We use the short term
\term{library@domain}{library domain} for any amount of knowledge, in particularly for the
knowledge in the global library. The term
\term{global@scientific@library}{global scientific library\/}%
\footnote{The term \term{GIL}{Global Intellectual Library, GIL,}
 is reserved for the project \url{http://sourceforge.net/gil/}.}
if used for a system managing the \term{global@library}{global library}.%
\footnote{Of course, we mean a network of systems functioning together.}
For brevity short term \term{scintific@library}{scintific library} is used.
\end{defn}

%
%

\section{Scientific document}

Here we talk about the main idea of knowledge representation in a space-time database.

\begin{defn}
An \term{abstract@document}{abstract document} is any information (knowledge)
represented in some way: on paper, photo-film, computer file etc.
As a rule, an \liketerm{abstract@document}{abstract document} possesses the structure
and is composed of a set of \term{abstract@elementary@documents}{elementary documents}:
plain text, mathematics, schemes and diagrams, graphics and other construction.
The main pourpose of the document, especially as regards scientific one,
is to pass (to give) to a reader a certain amount of knowledge, and do it most adequate.
Therefore an \termlink{abstract@document}{abstract document} can be characterized,
from system point of view, by the following properties:
\begin{itemize}
 \item[--] having the unambiguous sense (semantics)%
    \footnote{More exactly a document is sopposed to be precise and one-valued
        even in case when something is implicitly subtended, semantics include it as well,
        --- several meanings are the set of meanings. In any case document must
        be evaluated to the same for any attentive reader.};
 \item[--] use of common (standard)
        \termlink{abstract@elementary@documents}{elementary document} types
        to be recognized by all readers.
\end{itemize}
As a corollary, we see that documents are formed by simple and standard
\termlink{abstract@elementary@documents}{elementary documents}
using a small collection of simple rules of structuring.
\end{defn}

Let us take the following definition of the computer document representation.

\begin{defn}
\term{computer@document}{Computer document} is any information (knowledge) ---
representing an \termlink{abstract@document}{abstract document} ---
designed for reading by a reader or processing by an automated system and
able to be copied and trnsmitted (over communication networks).
\end{defn}

Following this definition we base our
\term{GIL@formalism}{formalism of the scientific library}
on a special definition of a \termlink{GIL@document}{document}, which
properties are different form that of computer document in usual sense.

\begin{defn}
To explain this let us observe that the most objects in an information system environment
have the following fundamental behavior:
\begin{itemize}
 \item[--] An object is localized somehow in the space
    (for example, it can be a row in a database table,
    identified by a key);
 \item[--] It has or can have properties that called \term{eent@attribute}{attributes}
    (or slots) of the object.
 \item[--] It is borned and exsts \term{eent@evolve}{evolving} until death.
    Object's evolution is usually change of its attributes.
    An attribute value can be reconstructed for now or for any moment in the past,
    if we track object's history.
\end{itemize}
We call an object with such behavior an \term{evolving@entity}{evolving entity}.
\end{defn}

A document in a usual sense is also a kind of information system entity evolving,
for example, in process of editing. We pay strong attention to document editing issues
from the global scientific workspace point of view. To our opinion, a document aquire value
from the start of editing process, long before the publishing point.

\begin{defn}
A \term{GIL@document}{document} of the scientific library is a permanent and persistend
object corresponding to {\bf a certain, given once and for all, non-changeable}
information (knowledge).
\end{defn}

This is formalized in the following definition.

\begin{defn}
\term{permanent@entity}{Permanent entity} is an object which is not considered to have history.
A \liketerm{permanent@entity}{permanent entity} is eternal and constant, more rigorously,
living in the generalized, \term{informational@spacetime}{informational space-time}.
\end{defn}

A \termlink{permanent@entity}{permanent entity} may be attached to
an \termlink{evolving@entity}{evolving entity} in some way, however,
the evolution, change of the document is not considered as evolution or dynamics in usual,
physical sense. Further, to formalize if we introduce the notion of
\term{development@of@knowledge}{information development} which means, first,
that new information grows from the ``existing'' information and, second, that
the new is connected to the preceding, not replaces it.
For insight see physical models \ref{Physical@models} below.

An abstract example of \termlink{permanent@entity}{permanent entity} is
an identifier, generated randomly at time $t$, if it is unique.
Another example, which is most important for us, is a text which state is fixed
at time $t$, say, Bethoven's Sonata \No\,1. Principically, we may assume that
the contents of this text is uniquelly defined even there are no media ---
somewhere another copy of the text would exist.

From this point we put into base of consideration the permanent model of scientific document.
This approach leads to new knowledge representation and retreival techniques.
For some imformation processing scenario it gives good performance thank to different way
of the knowledge representation.

\section{Physical models. Space-time databases physical origin}
\label{Physical@models}

Let us consider an exmple of mechanical system: a single mass point moving in a force field
(no matter what).
\term{}{Dynamical system} is a law describing the evolution of a system,
given by a differential equation, transformation group or stochastical process.
From the classical mechanics point of view the motion is described by
a \termlink{dynamical@system}{dynamical system},
representing the evolution of system's state --- in our case --- point $x(t)$ in the phase spaces
given by coordinates $q$ and impulse $p$ of the point.
Evolution of a state is described by $x(t)$ expressing the denendance on time~$t$.
The phase space is set ${\Set{R}^3 \times \Set{R}^3}$.

Relativistic theory consider the \term{world@curve}{world curve}
which is the full history of the mass point,
represented by the curve $(t, q(t))$.%
\footnote{Here the reader may be confused by the fact, that the state seems to be defined by
 the coordinate, but the impulse contained implicitly in the model as $m\dot q(t)$.}
If the space-time is supposed to be non-curved (having simple topology) then
the \term{spacetime}{space-time} is given by the product $\Set{R}^4 = \Set{R} \times \Set{R}^3$
of the time axis and the space in each inertial system.

Suppose that an observer makes sequential observation of the mass point states.
Each observation happens and correspond to ``a time'' and locates the mass at``a point in the space''.
We draw quatation marks to indicate that everything depends on inertial system coordinate system.
At the same time from geometrical point of view the construction is simplier:
the \term{world@curve}{world curve} is just a curve on the
\term{space@time}{space-time} manifold and observations correspond
to points on this curve.

Now returning to the global library consider an abstract document
evolving (\termlink{development@of@knowledge}{developing}) in editing process.
We consider any change of the document as birth of a new revised document.
It is natural to index it by the time of the revision and
a \liketerm{pseudo@space@coordinate}{pseudo-space coordinate}, say, a unique random identifier
mentioned above.%
\footnote{It would make sense to complement it by the space coordinate, though it
would be needed only to estimate the simultaneity of events happened at opposite sides
of the Galaxy. Observe however that the concept of
\termlink{development@of@knowledge}{knowledge developing} is less sensible to the
\liketerm{simultaneity@problem}{simultaneity problem}.}.
Thus, the sequence of the revised documents can be called the
\term{document@wordl@curve}{world curve of the document}.

Now consider the measurements of mass point state as documents of the global
knowledge base, stating that the mass is at point $q(t)$ at time~$t$.
Assume that our information about the mass is restricted by these documents
(mass is enfluenced by external factors). Then the document pseudo-curve interpolate
the real \term{world@curve}{world curve} of the mass.

Notice that the sequence of documents is discrete. Thus, we have no information on how
the mass evolve outside that world points. From mechanical point of view we just
approximate the continuous \term{world@curve}{world curve}, but as regards information
we just know that at some times $t_k$ we observe something at point $q(t_k)$, but
we cannot be sure that we see the same object! This example demonstrates the
relativity of such concepts as existence and evolution of information.
Indeed, if we deal with quantum theory object, say, electron then the relations
between measurements become non-trivial.

\section{Formalism of the global scientific library}

The purpose of our investigation is to show how one can formalize the space-time
database environment providing the global scientific knowledge representation environment and
the universal scientific workspace. We use the term \term{formalism}{formalism} or
\term{formal@system}{formal system} for
a system of formal meta-concepts logical constructions describing the matter.
A~\term{model}{model} is a system of real or imaginary objects and processes demonstrating
the structure and behavior satisfying a \termlink{formal@system}{formal system}.
In mathematics any system is studied as abstract object existing in world of abstract
mathematical constructions based on a fundamental mathematical formalism such as
Cermello--Frenkel axiomatics for Kantor set theory. Thus, a mathematical model of
a phenomenon deals with an \liketerm{image@of@phenomenonphe}{image} of the phenomenon,
not the ``real'' object.

Some features of the global knowledge environment concept are to be considered in
the real world context. Namely, some objects like \termlink{document}{document}
are provided with global meaning and existence independent on formalized view.
This leads to certain mixture of formal and physical models.
Roughly speaking, we will use mathematical language to speak about the formal constructions,
though they refer to absolute physical objects such as
the \term{knowledge@graph}{knowledge graph}.

To be precise we follow the following assumption. The
\term{GIL@Formalizm}{formalism of the scientific library (formalism)} extends
the mathematical formal language by adding several cathegories of symbols referring
the objects from the physical context. These \term{external@objects}{external objects}
are constant terms of the language. The main cathegories are
\termlink{permanent@entity}{permanent} and \termlink{evolving@entity}{evolving entities}
and relations between them, more exactly a restricted set of these objects described as follows.

\begin{axiom}
For all the permanent entities it is convenient to use one cathegory and one common term
considering any \termlink{permanent@entity}{permanent entity}
as a \termlink{document}{document}. Further we always use the term document.
A sequence of documents may be joined using \termlink{marked@doc@relation}{relations} marked by
the \termlink{marked@doc@relation@type}{relation type} which is also a document.
The sequence is joined or not joined, if yes then this fact is represented by
a \termlink{marked@doc@relation@instance}{relation instance}, a document as well.
\end{axiom}

\begin{axiom}
As for the cathegory of evolving entities gathering all entities describing objects
evolving in the real world, a common simplified formal representaion will be used.
Namely, the state an evolving entity is represented by attributes --- a kind of marked
binary relations involving both evolving entities and documents.
\end{axiom}

In our case evolving entities are pirmarily used for real workld objects managing the
scientific library as well as actors observing andinfluencing the library via the
managing system.

\medskip

Now we consider a preliminary model of the global knowledge representation which helps
to understand where the concept is originates and how it works.

Consider again an abstract document: a book or an article.
An important feature of any document is structuring --- the document is composed
from parts, each part has its own sructure and so on. Moreover each part, im most cases,
has independent syntactical value and can be conseder as smaller document.

For a document in usual sense we distinguish the graphical and logical structure.
The \term{doc@graphical@structure}{graphical structure} is a way of representing document
on paper, screen or other readable media. Obviously the logical structure is of most importance
to understand the document (document semantics), and the graphical structure is derivative
of the logical. This principle is explicitly used in markup languages TeX, HTML, XML etc.
Furthermore,
a document may be organized non-trivially, though the logical structure is expressed by
elementary, mostly hierarchial, construction like list, dictionary (map) etc.
If some information is represented in different way then it is probably too special to fit
in the common concept of document. However, the basic ``bricks'' we usually operate are
sufficient to cover $99.9\%$ of constructions we could imagine, the remainder can be
classified as elementary document (``black boxes'') so far.

\subsection{Document hierarchy.}

So, the structure of the document is \term{hierarchy}{hierarchial}%
\footnote{Of course, hierarchy is just one of the approaches to data representation.
 For example, widely used relational database phylosophy is orthogonal in some sense
 to hierarchial representation, as it deals with graph that is rarely a tree,
 objects are strongly cross-referenced. As regards the knowledge graph, hierarchy is just
 one possible behavior, but most important to document structuring model.}
: document is composed by
first-level parts. These parts are \term{child}{childs} of the initial document,
which is called \term{container}{container}. Thus documents can be viewed as nodes
forming a \term{tree}{tree}. This approach allows to consider all documents as nodes
of potentially infinite graph (the \termlink{knowledge@graph}{knowledge graph}).
A document can be reconstructed from the graph by tracking all descendants.

\subsection{Local document model.}

\begin{axiom} 
A document as a node of an hierarchy satisfies the following conditions:
\begin{itemize}
 \item[--]
    Document can be empty.
 \item[--]
    Document can be associated with an elementry content having special semantics:
    integer, real and complex numbers, string, graphical images etc.
 \item[--]
    Document can be associated with a \term{container@structure}{container structure}
    --- a way of organizing child nodes.
 \item[--]
    The following container structures are used:
   \termlink{container@dictionary}{dictionary} (\termlink{container@dictionary}{map}),
   \termlink{container@list}{list},
   \termlink{container@set}{set}.
 \item[--]
    \term{container@dictionary}{Dictionary} or (\liketerm{container@dictionary}{map})
    associated with a document $d_1$ is a map $m(d1) \Maps d_2 \mapsto d_3$,
    where $d_2$ called a \term{dictionary@key}{key},
    $d_1$, $d_2$, $d_3$ are documents.
    An \term{dictionary@element}{element of the dictionary}, $m(d_1)(d_2)$
    is called the $d_2$-\term{attribute@for@the@key}{attribute} of the document~$d_1$.
 \item[--]
    \term{container@list}{List} associated with a document $d_1$ contains
    ordered list of child documents which call \term{list@element}{elements of the list}.
    There are no principle obstacles for a list to be infinite.
 \item[--]
    \term{container@set}{Set} associated with a document $d_1$ contains
    finite or countable setordered list of child documents which call \term{list@element}{elements of the list}.
 \item[--]
    There are not principle obstacles for
    a \termlink{container@list}{list} or a \termlink{container@set}{set} to be infinite.
 \item[--]
    Container structures may be combined in the following way:
    Any node is a dictionary. However other pairs, where both elements are not a dictionary,
    are forbidden.
\end{itemize}
\end{axiom}

\begin{figure}[th]
 \label{kg_1}
    \includegraphics[-25mm,-33mm][50mm,10mm]{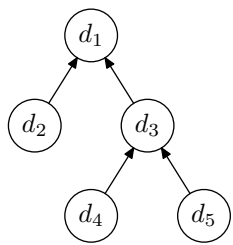}
    \caption{Simplified document structure}
\end{figure}

\subsection{Example of document life cycle.}
Here we consider an example of document life cycle --- how it is defined according to
the principle of permanent evaluation and influenced by the editing process.
Consider a document $d_1$ which is structured as shown on the figure~\ref{kg_1}:
The full content of the document $d_1$ is everything that is aggregated in it:
children, children of the children and so on. A document can enter different kinds of
relations but there is a restricted set of relations which, roughly speaking,
\term{evaluate@document}{evaluate} the document --- define its properties those change
would imply change of the document meaning (this process we call
 \term{development@of@document}{development} to distinguish with evolution
 applied to real world-like objects).

Any document change is related to an actor that performed it and can be considered as
a special kind of \term{revision@object}{revision object} associated with a time and a place.
An idea how to keep the permanent evaluation is well-known. We just produce branches
according to editing process. What is interesting is how to do it globally in a potentially
infinite world of knowledge both in ``horizontal'' (complexity, network) and
``vertical'' (hierarchy) direction?

\begin{figure}[th]
 \label{kg_2}
    \includegraphics[-35mm,-33mm][120mm,10mm]{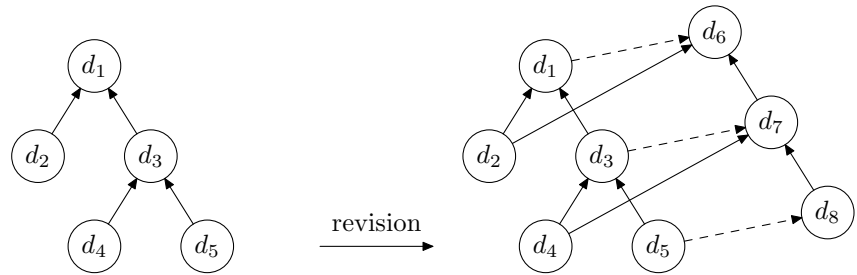}
    \caption{Document revision scenario}
\end{figure}

An idea is to track the revision everywhere where it can affect. Namely, a revision of a document
(which is a node of the knowledge graph) produces new revised documents for this document and
all the ancestors of it. In other words, if somebody (an \term{investigator}{investigator})
is importing a text and starting to edit he is supplied with {\em his own} copy od the text,
though it of course connected to the original document.

\begin{axiom}
This is to realize principle ``\term{pri@k@for@all}{Knowledge for people}'' ---
any reader can access to the knowledge system of mankind and develop it.%
\footnote{It would be interesting to descuss security and privacy issues here.
 Knowledge created (sometimes randomly) in a universal environment need additional
 instruments to control privacy and reliability, the same way like serious
 networked application overcome the Internet democracy.}
\end{axiom}

Using such kind of revisions expressed by knowledge development allows to exploit
different technological principles of managing data. Consider two
\termlink{investigator}{investigators} working with different domains:
each \term{GIL@domain}{domain} is a materialized part of the abstract global
knowledge system (associated with a workstation, a server or a storage).
Suppose they are editing the same document, for example, a medical health history of a patient
(`same' means branched from the same document --- node of the knowledge graph).
Let they enter different observation, which are probably different, but may be concurrent.
The problem of doing work in a client-server or distributed environment is rather non-trivial.
Actions are to be performed by specially designed API, put in transactions, monitored and logged,
checked for user rights and so on. Our case anything is considered as new knowledge just to be
added to the knowledge graph. More exactly, our
\term{knowledge@approach}{knowledge approach to information management} assumes that
we distinguish two tasks: 1)~processing document and 2)~managing document status.
The first deals with document structure, relations, non-trivial context (e.\,g.\ mathematics,
physical measurements, graphical design etc.), but the second concerns resolving revision status,
privacy and security issues. In our examples, to record the information is more important process
(one investigator would be an anaesthesist who has several seconds to do that)
than tracking observation branches and resolving status, which can be procedded later.

Thus, our two investigators just separately develop a copy of a document, which is a revision
of the original document. How they cooperate? Information synchronization is rather simple here.
Two domains can be joined any time. We can take new documents from one and to attach to another,
and no collisions happen since editing (development processes) produces new documents, having
different identification.

\begin{axiom}
This can be formulated as ``\term{pri@knowledge@growth}{Principle of knowledge grows}'':
abstractly the knowledge can only grow, anything {\em known} (at or from some point of the
 space-time) cannot be canceled or reversed.
\end{axiom}

Of corse, one may argue that any information system behaving like that will collapse when
information media cannot store more data which is never erased. Evidently, here we follow
the general idea of referencing and garbage collection. This idea is realized in some
programming languages like Java or Python operating with references to objects.
An object is holded in memory until reference count reaches zero, i.\,e.\ nobody needs this
object more and, what is important, nobody don't {\em know} more about this object and
cannot refer it. In a simmilar manner if a document is not needed --- neither readed
nor investigated more --- it can be erased. If a document is refered and blocked by
a reader or investigator we say that it is \term{observe}{observed}.
Some restrictions arise concerning collecting garbage. For example, if an object is a child
of an observed document it should be blocked as well and so on. This restrictions need predicate
calculus to be expressed.

Sure mentioned aspects concerning knowledge life cycle are not exhaustive but, we hope,
sufficient to understand the main ideas. We return to this discussion further.

\subsection{Modeling relations.}

\begin{defn}
A \term{relation}{relation} between elements of a set $M$ is a map
$r \Maps M^{\times n} \to \{\mathrm{True},\mathrm{False}\}$.
The relation can be equivalently represented by a set of sequences $(d_1,\dots,d_n)$
of elements, such that $r(d_1,\dots,d_n) = \mathrm{True}$, we say that
{\it these elements are in relation~$m$}.
\end{defn}

\begin{defn}
We call a \term{marked@doc@relation@instance}{marked relation instance}
(or a \liketerm{marked@doc@relation@instance}{link}) corresponding to a
\term{marked@doc@relation@type}{relation type} $m$ a finite sequence of documents
$(m,d_1,\dots,d_n)$.

From mathematical point of view we may consider all the relation instances as
elements of the set wich called \term{marked@doc@relation}{relation} of type given by~$m$.
\end{defn}

We reduce this notion to one ternary relation.

Indeed, the link can be viewed as an instance of {\em one universal} $(n+1)$-ary relation
instance $(d_0,d_1,\dots,d_n)$ where the beginning element plays role of relation type.
Furthermore, $(n+1)$-ary can be reduced to the marked binary relation as follows.
The relation is modeled by a \termlink{container@dictionary}{dictionary} instance
representing the map $k \mapsto d_k$, $k = 0,1,\dots,n$.

Observer that a \termlink{container@dictionary}{dictionary} as we has defined it
is a kind of ternary relation between container, key and item. However we cannot
go this way as we occupy the ternary relation just for representing dictionaries
and loose universality. We use the following solution%
\footnote{Other containers are build the same manner.}.
Let us introduce the constant document named `IsADictionaryAnchorOf'
marking the binary relation $(\mathrm{IsADictionaryAnchorOf},\cdot,\cdot)$ attaching
to a document $d_1$ no or one document $d_2$\ \,%
\footnote{By instance $(\mathrm{IsADictionaryAnchorOf},d_2,d_1)$}
called \liketerm{dictionary@anchor}{dictionary anchor}
of special meaning: a relation instance $(k_j,d_j,d_2)$ says that $d_j$ is the $k_j$-attribute
of~$d_1$.

\section{Semantics}

The key idea of the global knowledge environment is rather evident: to build complex
things with help of simple elements. A delicate feature is a way of understanfing this
common principle. The universal knowledge graph introduced above, based on plain node set
and the universal ternary relation is just an example of a simple and fundamental construction.
Other well-known examples are: representing mathematics 
formal logics on paper as a text --- a finite sequence of letters and symbols,
and formal models of algorithms like Turing machine (cf. Church-Turing thesis).
These constructions can be materialized in the real world without loss of meaning,
and different ways can be used: latin capital `A' drawn with any font in a mathematical text
means the same symbol, which is distinguished from `B'. Our knowledge graph can be also
represented in different ways: in computer format as described above 
or elsewise (say, in XML),
on paper or even as metallic construction with labelled balls and segments.

It is important that the primitive structure 
of the knowledge graph is understandable to any,
possibly non-human, reader able to perceive a construction like graph. 
This is a minimal requirement to make exchange of knowledge possible. 
Though a reader would take a long time to reconstruct it.

\begin{defn}
We call a real or imaginary process of interaction between the knowledge graph
and a being (creature) unfamiliar with it the 
\term{nonhuman@reader@test}{non-human reader test}.
\end{defn}

Thus the knowledge graph can be considered as a material, physical object.
This is explained as follows. The knowledge graph understood as described above 
need no additional semantics to deal with. Anything that does not concern
nodes and the relation is outside the concept. At the same time the knowledge graph
supporting by a knowledge management system must be outlined according to
it natural meaning.

To improve the understanding we must explain to the reader more complex constructions 
step by step. For doing this we introduce special semantical extensions called
\termlink{micromodel}{micro-models}, which allow to build advanced constructions 
on the base of the knowledge graph: 
hierarchy, revision, inheritance, metrics and so on.

\begin{defn}
A \termlink{micromodel}{micro-model} of the scientific library is a formal
requirement and semantic system to be applied to the knowledge graph
(more exactly, to a domain). A micro-model is a general notion, it is
represented by a document and included to the set of domain micro-models.
A micro-model attaches extra human semantics as well as constraints, constants etc.
Formally a micro-model is a part of the knowledge graph (constants and
fundamental relations) and a collection of rules in a human logics
(predicate calcules) involving both documents and human abstract constructions
introduced in micro-models.
\end{defn}




\end{document}